\documentclass[%
 reprint,
 amsmath,amssymb,
 aps,
prb,
showkeys
]{revtex4-2}

\usepackage{graphicx}
\usepackage{dcolumn}
\usepackage{bm}

\usepackage{mathrsfs}
\usepackage{float}      
\usepackage[table]{xcolor}      
 \usepackage{amsmath}
 \usepackage{amssymb}
 \usepackage{amsfonts}
\usepackage{multirow}
\usepackage{makecell}

\usepackage{xcolor}

\usepackage{hyperref}
\hypersetup{
    colorlinks=true,
    linkcolor=blue,
    citecolor=blue,
    urlcolor=blue
}

\usepackage[nameinlink,capitalize]{cleveref}

\crefname{equation}{Eq.}{Eqs.}
\Crefname{equation}{Eq.}{Eqs.}
\crefname{figure}{Fig.}{Figs.}
\Crefname{figure}{Fig.}{Figs.}
  
\begin{document}

\graphicspath{ {./} }
\preprint{APS/123-QED}

\title{Thermal entanglement and frustration: Probing quantum resources through concurrence and stabilizer structure}

\author{R. Pourkhodabakshi}
\email{rpourkhodaba@mun.ca}
\author{F. Dominie}
\author{D. Gajera}
\author{S.~H.~Curnoe}%
 \email{curnoe@mun.ca}
\affiliation{%
 Department of Physics and Physical Oceanography, Memorial University of Newfoundland,
St. John's, Newfoundland \& Labrador, Canada A1B 3X7
}%

\date{\today}

\begin{abstract}
In a frustrated spin system, highly entangled eigenstates can form a separable thermal mixture. We study this distinction for four spin-$1/2$ moments on a tetrahedron, the elementary unit of the pyrochlore lattice, with the general four-parameter exchange Hamiltonian. Tetrahedral symmetry allows a restricted search over thermal-state decompositions to be carried out by linear optimization, yielding an upper bound on multipartite concurrence and explicit separable decompositions where this bound vanishes. At low temperature, the concurrence maps show extended separable regions near the all-in--all-out limit, whereas suppression near the spin-ice point is narrowly localized. Heating broadens the latter region as nearby
multiplets are repopulated. The optimization identifies fully separable thermal states even when the eigenstate-averaged concurrence remains large. Negativity independently confirms finite-temperature entanglement in selected coupling regions. A mixed-state stabilizer R\'enyi diagnostic, used to explore ``quantum magic," 
can sometimes increase
with temperature. These results connect  entanglement of the thermal state to 
symmetry multiplets, and provide a practical basis for identifying frustrated spin configurations for studies of quantum resources beyond the ground-state limit.

\end{abstract}

    \keywords{Multipartite concurrence, thermal entanglement,
geometric frustration, spin systems, quantum magic}
\maketitle



\section{Introduction}

Entanglement describes quantum correlations that cannot be reproduced by a 
mixture of independent spin states~\cite{Horodecki2009,Amico2008}. In many-body systems, it provides a way to examine collective behaviour, including changes near quantum phase transitions~\cite{Osterloh2002,OsborneNielsen2002}. Its role in teleportation and measurement-based quantum computation also connects these correlations to information-processing tasks~\cite{Bennett1993,RaussendorfBriegel2001}.

Entanglement alone, however, does not determine computational power. Clifford circuits can generate highly entangled stabilizer states from computational-basis states while remaining efficiently simulable classically~\cite{Gottesman1998,AaronsonGottesman2004}. Quantum magic, or non-stabilizerness, describes the resource beyond mixtures of stabilizer states~\cite{Veitch2014}. Suitable magic states supply the non-Clifford operations needed for universal computation~\cite{BravyiKitaev2005,HowardCampbell2017}, while stabilizer decompositions relate this resource to classical simulation costs~\cite{Bravyi2019}. Studies of many-body magic thus provide a complementary perspective on interacting spin systems~\cite{LiuWinter2020,White2021,Sarkar2020}.

For a frustrated magnet, a natural question is which exchange couplings favour these resources and whether they persist at nonzero temperature. Thermal occupation can suppress entanglement or generate it by populating entangled excited states~\cite{Arnesen2001,Gunlycke2001}. The entanglement of individual eigenstates is therefore insufficient; the thermal density itself must be examined.

Pyrochlore magnets such as $\text{Tb}_2\text{Ti}_2\text{O}_7$ or $\text{Pr}_2\text{Zr}_2\text{O}_7$ are quantum spin systems that exhibit long range entanglement characteristic of a quantum spin liquid which is apparently due to the high symmetry and connectivity of the pyrochlore crystal structure.
In pyrochlore crystals, the spins are located at the vertices of corner-sharing tetrahedra, an arrangement known as geometric frustration.  
The space group symmetry is $Fd\bar{3}m$, which has 
face-centred cubic translations and an underlying octahedral point group symmetry which includes screw rotations which map tetrahdra of different orientations to each other. The most general model consistent with this symmetry  for nearest neighbour spin exchange interactions contains only four parameters ${\cal J}_i$ (exchange constants) ~\cite{Curnoe_physrevb_2008},
\begin{equation}
H_{\rm ex} = {\cal J}_1 X_1 + {\cal J}_2 X_2 + {\cal J}_3 X_3 + {\cal J}_4 X_4
\label{H:exchange}
\end{equation}
where
\begin{eqnarray}
X_1 & = & -\frac{1}{3} \sum_{\langle ij\rangle} J_{iz}J_{jz} \\
X_2 & = & -\frac{\sqrt{2}}{3} \sum_{\langle ij\rangle} [\Lambda_{ij}(J_{iz}
J_{j+} + J_{jz}J_{i+}) + {\rm h.c.}] \\
X_3  & = & \frac{1}{3} \sum_{\langle ij \rangle } (\Lambda_{ij}^{*}
J_{i+}  J_{j+} + {\rm h.c.} )  \\
X_4 & = & -\frac{1}{6} \sum_{\langle ij \rangle }(J_{i+} J_{j-}  + {\rm h.c.})
\end{eqnarray}
and where $\langle ij\rangle$ is a pair of nearest-neighbours, 
h.c. stands for ``Hermitian conjugate," $J_{\pm}= J_x\pm i J_y$, and $\Lambda_{ij}$ are phases that depend on the sites.
In these expressions the Cartesian axes are {\em local} axes, defined for each spin site 
on the vertices of a tetrahedron such that the local $z$-axis points from the centre of the tetrahedra to the spin site.  The local axes and the scheme for selecting them are described in \cref{sec:symm}; full details about the site-labeling, phases and local axes are given in  \cite{Curnoe_physrevb_2008}.   

The model given by \cref{H:exchange}  is valid for {\em any} angular momentum $\vec{J}$, in particular it is the most general model for spin-1/2.  In rare earth pyrochlore crystals,   $J$ is determined by Hund's rules, for example, $J=6$ for Tb$^{3+}$ or $J=8$ for Ho$^{3+}$. However, in general the $2J+1$-fold degeneracy is lifted by the crystal electric field (CEF) at the magnetic sites such that the lowest energy state is a singlet or one of three kinds of doublet \cite{Curnoe2018}.  In this paper we will focus on the cases where the CEF ground state is a doublet that is isomorphic to spin-1/2, that is, it transforms as a spinor.  Some examples where the CEF ground state transforms as a spin-$1/2$ spinor are Yb$_2$Ti$_2$O$_7$ and Er$_2$Ti$_2$O$_7$~\cite{Curnoe2018}.

In general the model (\cref{H:exchange}) can only be solved numerically, except for the special case when ${\cal J}_{2,3,4} = 0$. 
In this case, when ${\mathcal{J}}_1 >0$ the ground state configuration, known as the ``all-in-all-out" state, is one for which on every tetrahedron all the spins are either point toward or away from the centre of the tetrahedron.     When ${\mathcal{J}}_1 >1$ the ground state is the highly degenerate ``spin ice" state, which has two spins pointing into and two spins pointing out of each tetrahedron on the lattice (``two-in-two-out" configurations).   Spin ice states are classical and unentangled, but non-zero values of the other constants will lift the degeneracy of the spin ice manifold and produce entangled eigenstates of $H_{\rm ex}$.   However, the entanglement is expected to be evident only at very low temperature. Concurrence and entanglement have also been studied on a 16-site spin-$1/2$ pyrochlore cluster ~\cite{Chen2025-PhysRevB.111.014436}. 

A single tetrahedron is smallest sub-unit of the pyrochlore crystal that possesses the point group symmetry (omitting the screw rotations.  The general exchange Hamiltonian for this system is identical to (\cref{H:exchange}),
and eigenstates are similar to the full-lattice model; in particular the ground state for ${\mathcal{J}}_1 >0$ and ${\mathcal{J}}_{2,3,4} = 0$ is the set of 6-fold degenerate two-in-two-out states.   This small system can be easily solved for any values of the exchange constants ${\cal J}_i$. The degeneracies of the eigenstates are determined by symmetry \cite{Curnoe2007_PhysRevB_2007}: in general the 16-dimensional space splits into a singlet, 3 doublets and 3 triplets, all of which are entangled in general.  However, the entanglement for each individual eigenstate does not tell us much about the entanglement of the system, which is more realistically described as a thermal density arising due to interactions with an environment.  The thermal density  accounts for the degeneracies and the spectrum 
of all the states at finite temperature.  However, as we shall see, it is much more difficult to quantitatively evaluate the entanglement of a density than the individual pure states of which it is composed. 
This paper discusses and compares the results of different  quantitative approaches to evaluating entanglement of spins on a tetrahedron at finite temperature.


Evaluating the mixed-state concurrence requires an optimization over pure-state decompositions~\cite{WootersPhysRevLett.80.2245,carvelloPhysRevLett.93.230501,mintert2005}, whereas negativity offers a directly computable, though incomplete, entanglement test~\cite{VidalWerner2002}. Symmetry can simplify such optimizations~\cite{TerhalVollbrecht2000,VollbrechtWerner2001}. Here we use tetrahedral symmetry to obtain upper bounds on concurrence through constrained linear optimization, and compare them with negativity and a stabilizer R\'enyi diagnostic~\cite{Leone2022}. This reduces the computational cost of examining different couplings and temperatures, providing a starting point for identifying spin configurations that combine entanglement and useful magic beyond the ground-state limit.

\section{Quantifying entanglement}

Entanglement in a pure state can be quantified in different ways (see \cite{mintert2005, Gajera_2026} for a discussion).  For example, the $I$-concurrence measures the entanglement between two parts of an $N$-partite state,
$$C_I(\psi) = \sqrt{1 - \mbox{Tr}[\rho_i^2]},$$
where $\rho_i$ is the reduced density of 
$\rho = |\psi\rangle \langle \psi|$ 
obtained by taking the trace over one of the parts, while an often-used definition for concurrence takes into account all of the $I$-concurrences, 
\begin{equation}
C( \psi) = 2^{1 - \tfrac{N}{2}} \sqrt{(2^N - 2) - \sum_i \operatorname{Tr}(\rho_i^2)},
\label{N-concur}
\end{equation}
where $\rho_i$ are the reduced densities of 
resulting from all possible ways of partitioning the system into two parts.  Unlike the $I$-concurrence, this expression possesses the same symmetry as the state itself, and will therefore be most useful for our purposes. 

A mixed state takes the form
\begin{equation}
    \rho = \sum_i p_i |\psi_i\rangle \langle \psi_i|
\end{equation}
where $\sum_i p_i = 1$ and $p_i >0$.  Since the decomposition of $\rho$ into pure states is not unique \cite{hughston1993},
the generalization of \cref{N-concur} to densities  is obtained by finding the decomposition 
\begin{equation}
    \rho = \sum_j q_i |\phi_j\rangle \langle \phi_j|,
\end{equation}
that minimizes $\sum_j q_j C(\phi_j)$, that is
\begin{equation}
      C(\rho) =  \inf_{\{q_i,  \phi_i\}}\sum_j q_j C(\phi_j).
      \label{infimum}
\end{equation}

Related constructions of concurrence based on conjugations are discussed in \cite{UhlmannPhysRevA.62.032307}.
Given computational challenges of computing the infimum in (\ref{infimum}), the {\em negativity} is often evaluated instead.
For multi-partite systems, the negativity is usually taken to be the average of the quantities 
$$
N_i(\rho) =2  \sum |\lambda|,
$$
where $\lambda$ are the negative eigenvalues of $\rho_i^T$, a partial transpose of $\rho$. 
For two spins (or qubits), concurrence provides a quantitative description of entanglement~\cite{Hill-WootersPhysRevLett.78.5022,WootersPhysRevLett.80.2245}.
The negativity faithfully captures entanglement in pure states. For a pure  bipartite state of the form
$|\psi\rangle = \alpha|++\rangle + \beta|+-|\rangle + \gamma|-+\rangle + \gamma|--\rangle$ the concurrence is 
$C(\psi) = 2 |\alpha \delta - \beta\gamma|$ and the negativity is $N(\psi) =C(\psi)$. Positivity of the partial transpose is necessary for separability, but is not sufficient in general~\cite{HORODECKI19961}. In general, the negativity is non-zero for an $N$-partite pure entangled state, 
however, the negativity of a mixed state can be zero even when the concurrence is non-zero.  

\section{Symmetry-guided evaluation of concurrence}  
\label{sec:symm}
In highly symmetric quantum systems the Hamiltonian eigenstates are superpositions of 
symmetry related states, which tends to favour entanglement.  Symmetry also introduces degeneracies such that a mixed state formed from degenerate eigenstates will tend to be less entangled than the individual eigenstates. Symmetry has also been used to evaluate concurrence-based entanglement measures for isotropic states~\cite{rungtaPhysRevA.67.012307}. Our symmetry-guided evaluation of concurrence~\cite{Gajera_2026} is a method that exploits the symmetries in a system to find an upper bound on the infimum in \cref{infimum}.

In this article we examine spin states on a single tetrahedron, which has the point group symmetry $T_d$ as well as time reversal symmetry ${\cal K}$.  
The spins are located at the vertices with  site symmetry  $D_{3d}$, which has a set of three-fold rotations about the axis pointing from the centre of the tetrahedron to the site (the $C_3$ axis), three two-fold rotations about axes perpendicular to the $C_3$ axis (the $C_2$ axes), and a similar set of improper rotations including inversion.  According to our procedure \cite{Gajera_2026}, we define the local $z$ axes to be in the direction of the highest symmetry axes, which are the $C_3$ axes, while the local $y$ axes point in the direction of one the $C_2$ axes.  The local $x$-axes are selected to obey the right hand rule. These definitions align with those commonly used to model magnetic interactions in pyrochlores \cite{Curnoe_physrevb_2008}.
Expressing the spin states with respect to the local axes  allows us to easily determine the degeneracies of the Hamiltonian (\cref{H:exchange}) and block diagonalize it \cite{Curnoe2007_PhysRevB_2007}.  

We will assume that the spin state of the tetrahedron is in thermal equililbrium,
$$
\rho = \frac{1}{Z} \sum_i e^{-E_i/T} |\psi_i\rangle \langle \psi_i|
$$
where $|\psi_i\rangle$ are the eigenstates of the Hamiltonian with energies $E_i$ and $Z = \sum_i e^{-E_i/T} $ is the partition function.   
Taking into account the degeneracies in the spectrum, there are seven distinct energy eigenvalues $E_j$ each belonging to a set of eigenstates
$|\psi_j^{(k)}\rangle$, where $k$ indexes the degenerate states.  The concurrence $C_j$ of each state can be calculated using 
Eq.\ \ref{N-concur}; degenerate states have the same concurrence.

The thermal density can be written as 
\begin{equation}
\rho = \sum_{j=1}^7 p_j \rho _j \label{thermal2}
\end{equation}
where 
\begin{equation} 
\rho_j = \frac{1}{d_j} \sum_{k=1}^{d_j}  |\psi_j^{(k)}\rangle \langle \psi_j^{(k)}|
\label{thermal_rho}
\end{equation}
and $d_j$ is the degeneracy of the eigenvalue $E_j$.  The probabilities
\begin{equation}
p_j = \frac{d_j e^{-E_j/T}}{Z} 
\label{probs}
\end{equation}
can be calculated.  
According to the above, the densities $\rho_j$ are composed of   equally weighted terms 
of the form $|\psi^{(k)}_j\rangle\langle \psi^{(k)}_j|$ which are {\em partners} with each other, that is they are a complete set of states related to each other by the operations of the symmetry group.  This means that by construction, each $\rho_j$ is invariant under the symmetry group $T_d$.  

While in general the density matrix of a system with 4 spins has
$16^2 - 1$ independent matrix elements, 
the most general form of a density that is invariant under the symmetry group of the tetrahedron has only ten independent elements.  Thus, instead of expressing the density as a $16\times 16$ matrix, it can be written as an 11-element array, where the first three elements are the dependent diagonal elements (which are constrained to sum to one) and the remainder are a set of independent off-diagonal elements.  Each of the seven densities $\rho_j$ (\cref{thermal_rho}) can be expressed in this way.

In order to evaluate  the concurrence we tackle the problem of the infimum using the approach described in Ref. \cite{Gajera_2026}. Essentially, we must systematically construct a large number of densities  that respect the tetrahedral symmetry of our system and that are {\em explicitly unentangled.} Following the procedure described in \cite{Gajera_2026}, we constructed 67 non-redundant completely symmetric unentangled densities expressed as 11-element arrays.  Redundant densities - those  that can be expressed as a positive sum of other densities - are not included in the set of unentangled densities.  
The set of seven 11-component array representations of the  densities $\rho_i$ in \cref{thermal_rho} are not orthogonal to each other, but they are  independent.  To form a complete independent basis for this 11-component space, we select four of the 67 unentangled density arrays that are also independent and add them to our set, and label them as $\rho_{8-11}$. The density \cref{thermal2} can be written as 
$$
\rho = \sum_{j=1}^{11}
p_i\rho_i,$$
where $p_{1-7}$ are given by \cref{probs}
and $p_{8-11} = 0$.

The remaining unentangled densities are denoted
$\eta_{1-63}$, and each of these can be expressed as a linear combination of $\rho_{1-11}$, 
$$
\eta_l = \sum_{i=j}^{11} c_{il} \rho_j,
$$
where the constants $c_{il}$ can be calculated.

To address the infimum problem,
we seek a new decomposition of the thermal density 
of the form
\begin{eqnarray}
    \rho &=& \sum_{i=1}^{11} p_i' \rho_i
    + \sum_{l=1}^{63} q_l \eta_l \\
    & = & \sum_{i=1}^{11}
    \left[p_i' + \sum_{l=1}^{63} c_{il} q_l\right] \rho_i,
\end{eqnarray}
{\em i.e.}, $p_i = p_i' + \sum_{l=1}^{63} c_{il} q_l$, 
that minimizes the quantity
\begin{equation}
     \sum_i p_j'C_j,
\end{equation}
where $C_j = C(\psi_j)$ are the concurrences of the system eigenstates.  
This is a straight-forward linear optimization problem in the variables $q_{1-63}$ with constraints
\begin{eqnarray}
   0 & \leq &  q_l \leq 1  \\
    0 & \leq & p_i - \sum_{l=1}^{63} c_{il} q_l \leq 1.
\end{eqnarray}
The minimum thus obtained is denoted $C_s(\rho)$, and it is an upper limit of the concurrence $C(\rho)$
given by \cref{infimum}.

\begin{figure}[htbp]
    \centering
    \includegraphics[width=80mm]{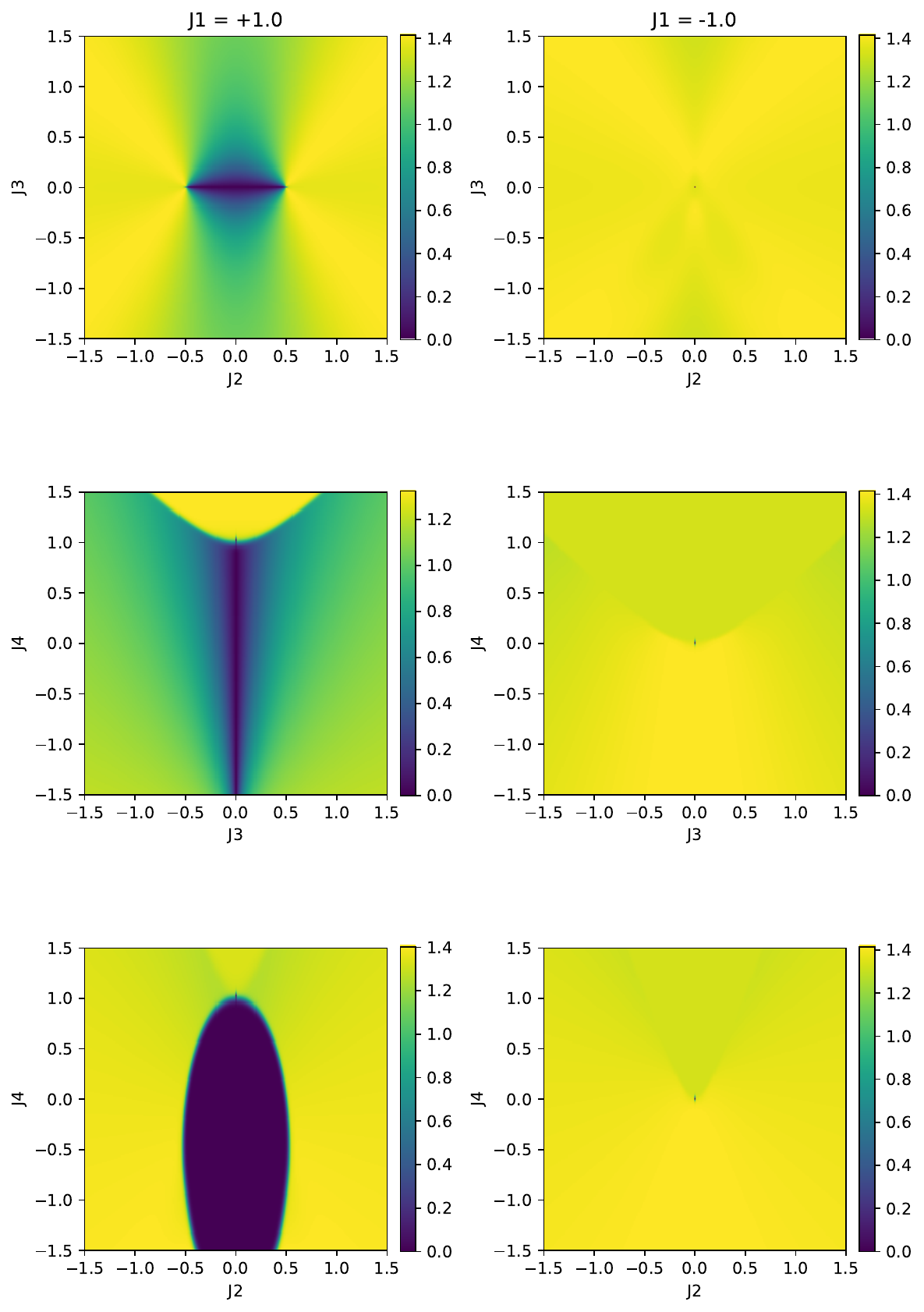}
    \caption{Optimized concurrence $C_s$ at $T=0.01$. Left column ${\mathcal{J}}_1=+1$, right column ${\mathcal{J}}_1=-1$; rows scan the $({\mathcal{J}}_2,{\mathcal{J}}_3)$, $({\mathcal{J}}_3,{\mathcal{J}}_4)$ and $({\mathcal{J}}_2,{\mathcal{J}}_4)$ planes with the third coupling set to zero.}
    \label{fig:concurrence_4}
\end{figure}

\section{Results \label{sec-results}}

\begin{figure}[htbp]
    \centering
    \includegraphics[width=80mm]{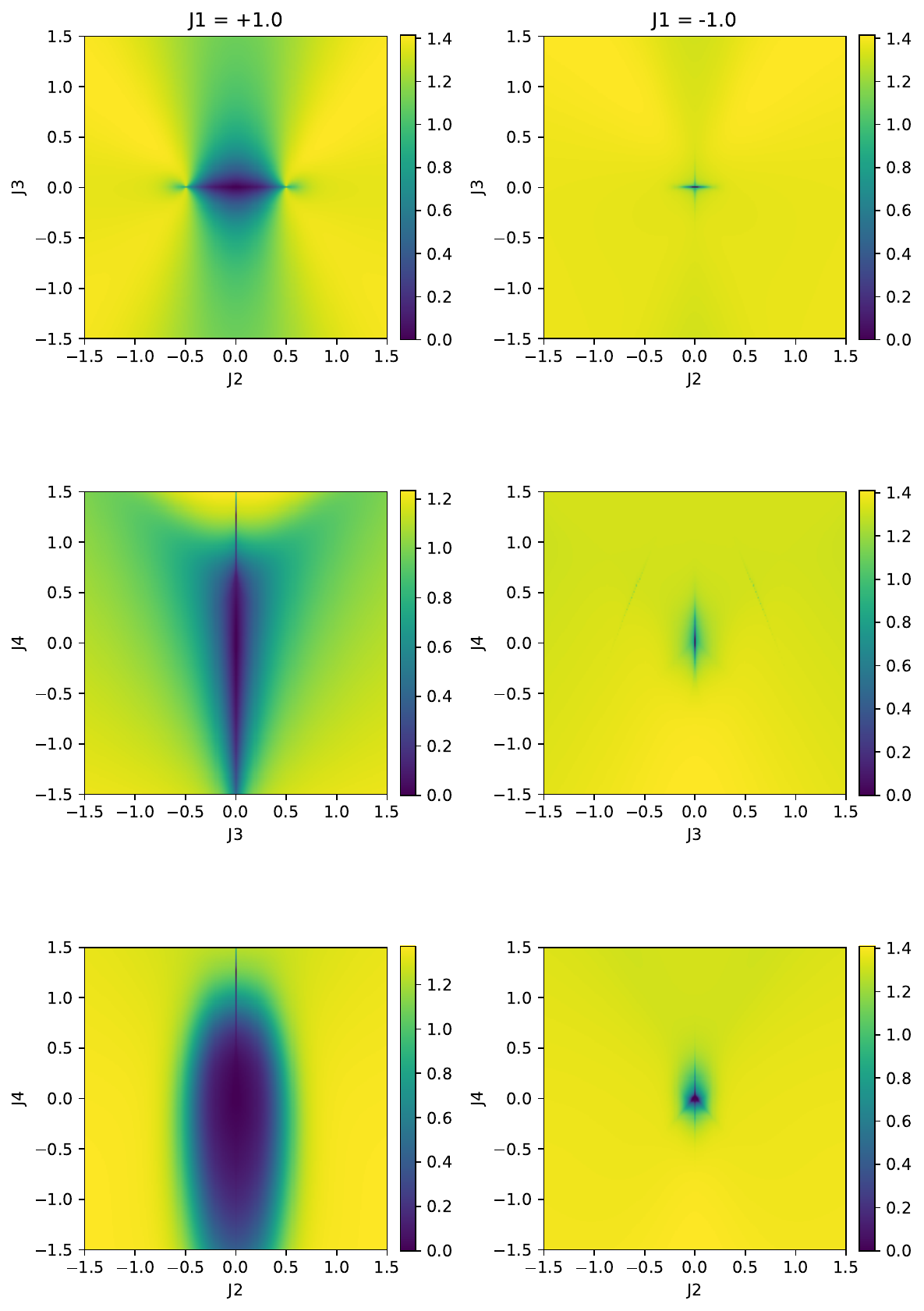}
    \caption{Optimized concurrence $C_s$ at $T=0.1$. Left column ${\mathcal{J}}_1=+1$, right column ${\mathcal{J}}_1=-1$; rows scan the $({\mathcal{J}}_2,{\mathcal{J}}_3)$, $({\mathcal{J}}_3,{\mathcal{J}}_4)$ and $({\mathcal{J}}_2,{\mathcal{J}}_4)$ planes with the third coupling set to zero.}
    \label{fig:concurrence_3}
\end{figure}

\begin{figure}[htbp]
    \centering
    \includegraphics[width=80mm]{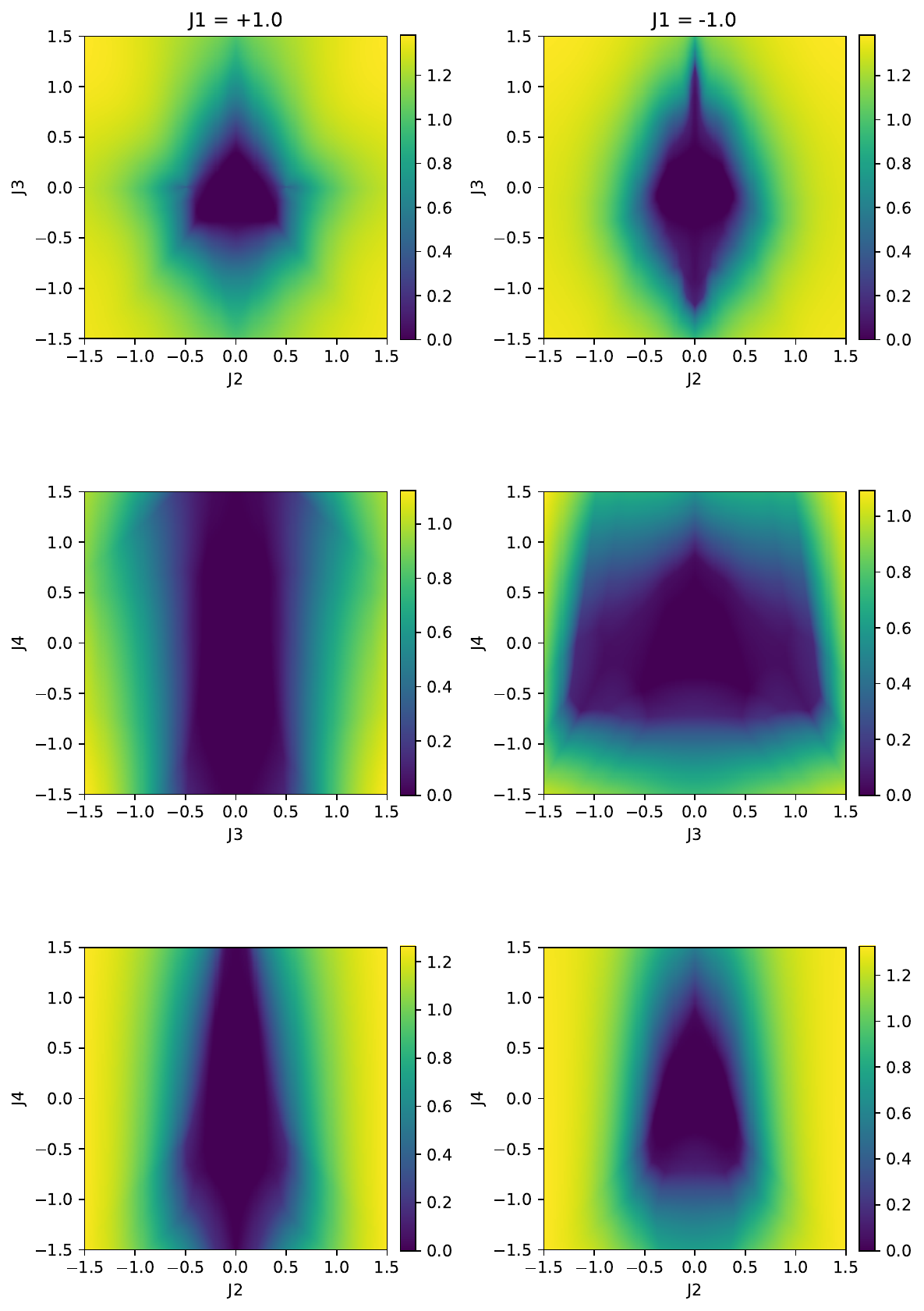}
    \caption{Optimized concurrence $C_s$ at $T=0.5$. Left column ${\mathcal{J}}_1=+1$, right column ${\mathcal{J}}_1=-1$; rows scan the $({\mathcal{J}}_2,{\mathcal{J}}_3)$, $({\mathcal{J}}_3,{\mathcal{J}}_4)$ and $({\mathcal{J}}_2,{\mathcal{J}}_4)$ planes with the third coupling set to zero.}
    \label{fig:concurrence_11}
\end{figure}

We map the optimized concurrence $C_s$ of the thermal state of a single tetrahedron
as a function of the exchange couplings, for the two signs of the dominant coupling
$\mathcal{J}_1=\pm1$ and for three temperatures $T=0.01,\,0.1,\,0.5$, shown in \cref{fig:concurrence_4,fig:concurrence_3,fig:concurrence_11}.
Each figure is a $3\times2$ array of heat maps: the left column fixes $\mathcal{J}_1=+1$ and the
right column $\mathcal{J}_1=-1$, while the three rows scan the $(\mathcal{J}_2,\mathcal{J}_3)$, $(\mathcal{J}_3,\mathcal{J}_4)$ and
$(\mathcal{J}_2,\mathcal{J}_4)$ planes with the remaining coupling held at zero. Each panel is auto-scaled
and carries its own colour bar, so colours should be read against the adjacent scale and
not compared directly between panels.
 
Throughout we use the multipartite concurrence measure
\cref{N-concur}
which for $N=4$ reduces to $C=\tfrac12\sqrt{14-\sum_i\operatorname{Tr}\rho_i^2}$, the sum
running over all $2^4-2=14$ reduced densities. The strongly entangled eigenstates saturate
near $\sum_i\operatorname{Tr}\rho_i^2=7$, giving $C(A_1)=C(T_{2\alpha})=\sqrt7/2\approx1.32$,
while the ceiling of the measure on four spins is $\tfrac12\sqrt{8.5}\approx1.46$. The bright
plateaus in every panel lie between these two values: wherever the ground state is a single
symmetry-entangled eigenstate the concurrence is large and only weakly parameter dependent.
The physics of interest is therefore located in the \emph{dark} regions, where $C_s$ is suppressed.

Two limiting ground states organize the whole picture. For $\mathcal{J}_1>0$ (left columns) the ground
state is the product ``all-in--all-out'' doublet $\{|{++++}\rangle,|{----}\rangle\}$, which is
separable and hence has $C_s=0$. For $\mathcal{J}_1<0$ (right columns) the lowest manifold is the
six-fold degenerate 2-in-2-out (spin-ice) set 
Each of these
states is a superposition of spin ice states and is \emph{individually} entangled, but at the fully degenerate point their equal mixture
can be re-expressed in terms of the separable symmetric densities $\eta_k$, so that $C_s\to0$
there as well. The two columns thus isolate the two distinct mechanisms our method is built to
distinguish: a genuinely separable ground state versus a degeneracy-reduced mixture of entangled
states.
 
\paragraph*{All-in--all-out ($\mathcal{J}_1=+1$).}
At the lowest temperature (\cref{fig:concurrence_4}, left column) the separable ground state
produces extended regions of \emph{exactly} vanishing concurrence, bounded by sharp
level-crossing curves. The shape of these null regions encodes which coupling entangles the
product state. In the $(\mathcal{J}_2,\mathcal{J}_3)$ plane the null region is a thin lens hugging the $\mathcal{J}_3=0$
line: any $\mathcal{J}_3\neq0$ mixes 
all-in--all-out) states with the entangled states
and immediately switches on the concurrence, whereas along $\mathcal{J}_3=0$ the product
state survives as a pure eigenstate (energy $-\mathcal{J}_2/2$) until a crossing near $|\mathcal{J}_2|\approx0.5$.
The $(\mathcal{J}_2,\mathcal{J}_4)$ panel shows the largest null region, a filled lobe reaching $\mathcal{J}_4\approx1$
and extending further for $\mathcal{J}_4<0$: along $\mathcal{J}_2=0$ the ground manifold remains a
symmetry-degenerate doublet/triplet 
whose mixture is still
separable, and only a finite $\mathcal{J}_2$ lifts that degeneracy into a non-reducible, entangled ground
state that lights up the map. In other words the boundary of the dark lobe is a locus where the
$\eta_k$ construction can no longer absorb the degenerate mixture.

\paragraph*{Spin ice ($\mathcal{J}_1=-1$).}
The right columns are essentially the mirror image: the map is bright almost everywhere and $C_s$
is suppressed only at the single frustrated point $\mathcal{J}_2=\mathcal{J}_3=\mathcal{J}_4=0$. This is the tetrahedral
analogue of the frustration-to-entanglement transition anticipated in the Introduction (and seen
earlier for the triangle \cite{Gajera_2026}): the 2-in-2-out degeneracy pins the concurrence to zero exactly at the
origin, but any perturbation splits the manifold and, because the constituents are themselves
entangled, the ground state jumps \emph{discontinuously} to a highly entangled configuration as
$T\to0$. The very low temperature panels (\cref{fig:concurrence_4}, right) show only a
point-like or faint cross-shaped null at the origin, consistent with this picture.

\begin{figure}[htbp]
    \centering
    \includegraphics[width=90mm]{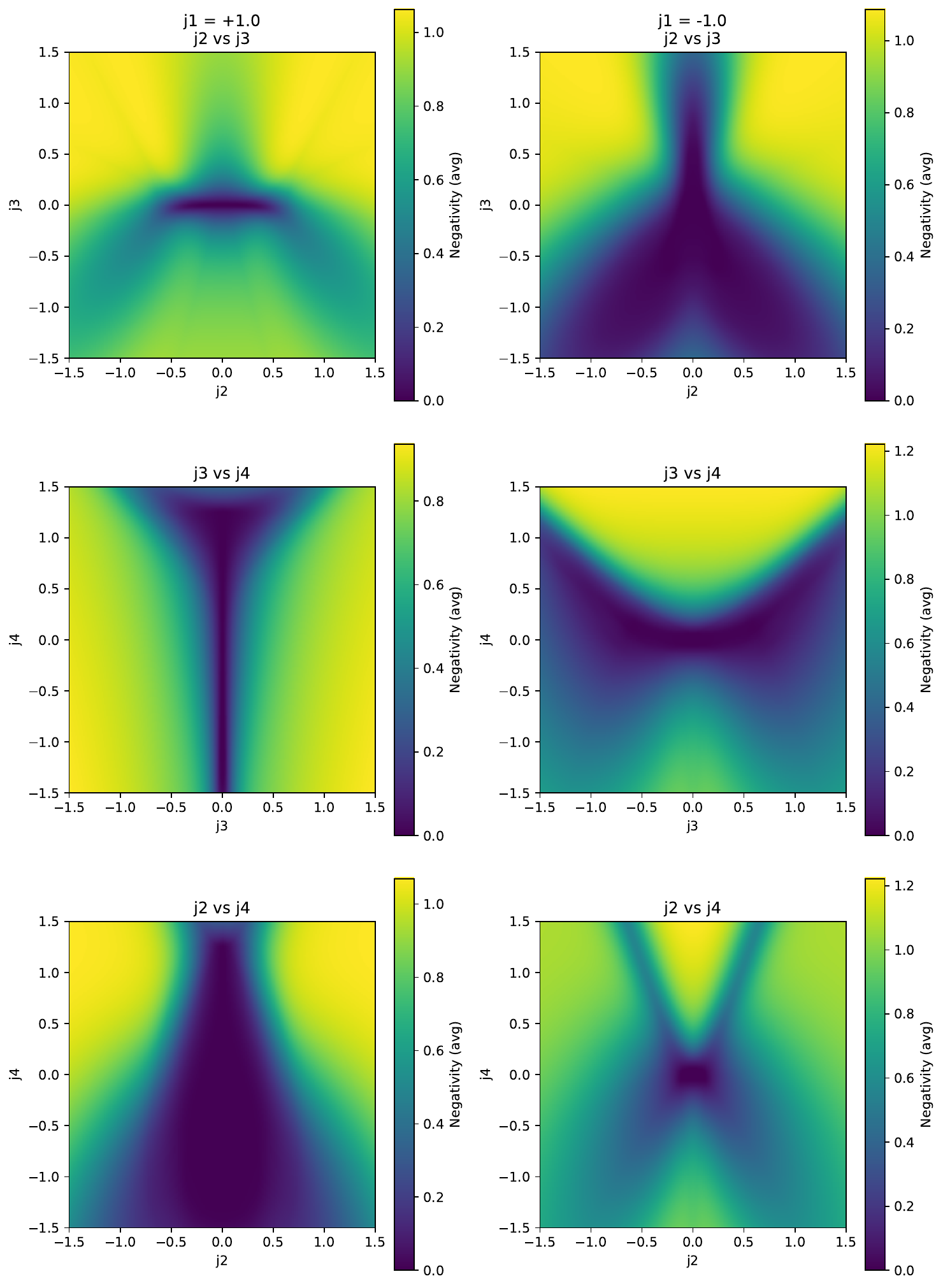}
    \caption{Average negativity $\bar N$ at $T=0.1$. Column/row layout identical to the concurrence figures (left ${\mathcal{J}}_1=+1$, right ${\mathcal{J}}_1=-1$; rows scan $({\mathcal{J}}_2,{\mathcal{J}}_3)$, $({\mathcal{J}}_3,{\mathcal{J}}_4)$, $({\mathcal{J}}_2,{\mathcal{J}}_4)$ with the third
coupling set to zero).}
    \label{fig:negativity_4}
\end{figure}

\paragraph*{Temperature dependence.}
Raising $T$ acts oppositely in the two columns. For $\mathcal{J}_1=+1$ the sharp crossing boundaries
soften into gradients and the null lobes shrink and partially fill in, because thermally populated
excited states contribute their own entanglement
(\cref{fig:concurrence_3,fig:concurrence_11}, left). For $\mathcal{J}_1=-1$ the trend is
reversed: the point-like null at $T=0.01$ swells into a finite dark region (the diamond-,
tent- and cup-shaped features at $T=0.5$, right column) of extent set roughly by
$|\mathcal{J}_a|\lesssim T$. Physically, once the temperature exceeds the small splitting of the spin-ice
manifold the six near-degenerate levels are repopulated with comparable weight, reconstituting the
separable degenerate mixture over an entire neighbourhood of the origin rather than at a single
point. The three temperatures together are thus a direct visualisation of the claim that this
transition can only be captured at finite $T$, with all of the low-lying, nearly degenerate states
retained in the density matrix.

\begin{figure}[htbp]
    \centering
    \includegraphics[width=80mm]{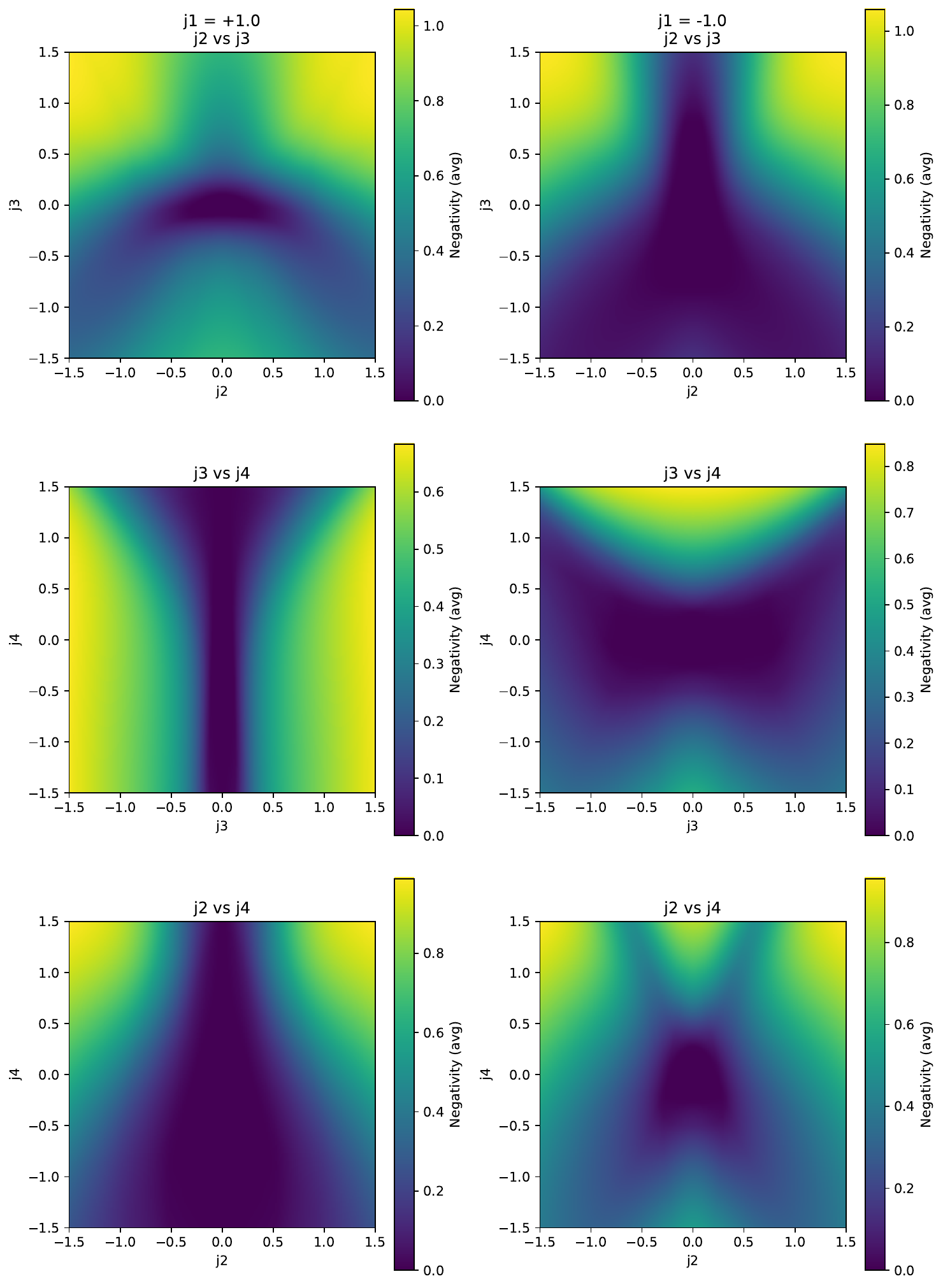}
    \caption{Average negativity $\bar N$ at $T=0.2$. Column/row layout identical to the concurrence figures (left ${\mathcal{J}}_1=+1$, right ${\mathcal{J}}_1=-1$; rows scan $({\mathcal{J}}_2,{\mathcal{J}}_3)$, $({\mathcal{J}}_3,{\mathcal{J}}_4)$, $({\mathcal{J}}_2,{\mathcal{J}}_4)$ with the third
coupling set to zero).}
    \label{fig:negativity_2}
\end{figure}


\subsection{Negativity}
\begin{figure}[htbp]
    \centering
    \includegraphics[width=80mm]{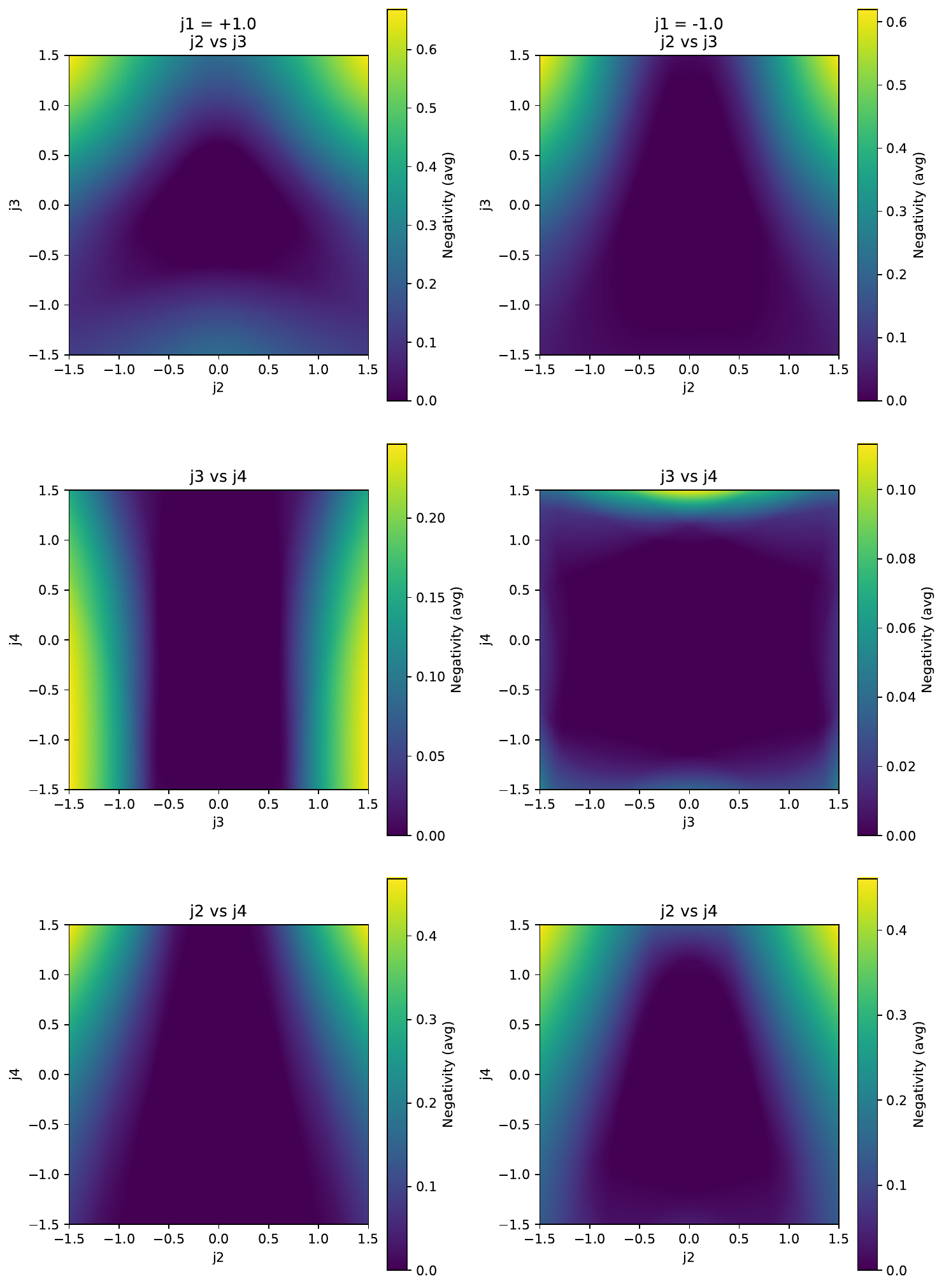}
    \caption{Average negativity $\bar N$ at $T=0.5$. Column/row layout identical to the concurrence figures (left ${\mathcal{J}}_1=+1$, right ${\mathcal{J}}_1=-1$; rows scan $({\mathcal{J}}_2,{\mathcal{J}}_3)$, $({\mathcal{J}}_3,{\mathcal{J}}_4)$, $({\mathcal{J}}_2,{\mathcal{J}}_4)$ with the third
coupling set to zero).}
    \label{fig:negativity_3}
\end{figure}

As an independent probe we compute the negativity of the same thermal states. For a bipartition
$A|\bar A$,
\begin{equation}
N(\rho)=\lVert\rho^{T_A}\rVert_1-1=2\sum_{\lambda_k<0}|\lambda_k|,
\label{nega-bipart}
\end{equation}
where the $\lambda_k$ are the eigenvalues of the partial transpose $\rho^{T_A}$; the factor of two
is chosen so that $N$ is faithful on pure bipartite states, $N(\Psi)=C(\Psi)$. The maps in
\cref{fig:negativity_4,fig:negativity_3} show the average $\bar N$ over the four $1|3$
cuts and the three $2|2$ cuts. Two features distinguish $\bar N$ from the optimized concurrence
$C_s$ at the outset: it is evaluated directly on $\rho$, with no convex-roof step and no symmetric
separable densities $\eta_k$ entering; and it witnesses entanglement through the positive
partial transpose (PPT) criterion rather than through the entanglement of formation.

It is useful to fix the scale on a pure eigenstate. For the $2$-in-$2$-out state $|A_1\rangle$ every
$1|3$ reduction is $\rho_1=\tfrac12\mathbb{1}$, so each $1|3$ cut saturates its maximum, $N=1$;
each $2|2$ cut has Schmidt amplitudes $\{1/\sqrt6,1/\sqrt6,2/\sqrt6\}$, giving
$N=(4/\sqrt6)^2-1=5/3$. The average is $\bar N(A_1)=\tfrac{1}{7}(4\cdot1+3\cdot\tfrac53)=\tfrac97
\approx1.29$, essentially equal to $C(A_1)=\sqrt7/2\approx1.32$. Thus on a single dominant eigenstate
the two measures agree; any disagreement in the maps must come from \emph{mixing}.

\paragraph*{Matched temperature (\cref{fig:negativity_4}, $T=0.1$).}
For ${\mathcal{J}}_1=+1$ (left column) $\bar N$ reproduces the same all-in--all-out (AIAO) null
structures seen in the $T=0.1$ concurrence map: the horizontal null lens along ${\mathcal{J}}_3=0$ in the
$({\mathcal{J}}_2,{\mathcal{J}}_3)$ plane, the vertical dark stripe at ${\mathcal{J}}_3=0$ in
$({\mathcal{J}}_3,{\mathcal{J}}_4)$, and the dark plume about ${\mathcal{J}}_2=0$ in $({\mathcal{J}}_2,{\mathcal{J}}_4)$. The
bright plateaus reach $\bar N\approx1.0$--$1.2$, comparable to the 
concurrence plateau; this is
expected, since the AIAO ground state 
is a genuine product state that both measures
report as separable. The negativity nulls are, however, softer-edged than the concurrence nulls: with
no $\eta_k$ absorption, $\bar N$ is lifted off zero as soon as excited entangled states acquire
thermal weight, so it decays smoothly rather than vanishing over a sharply bounded region.
\begin{figure}[htbp]
    \centering
    \includegraphics[width=80mm]{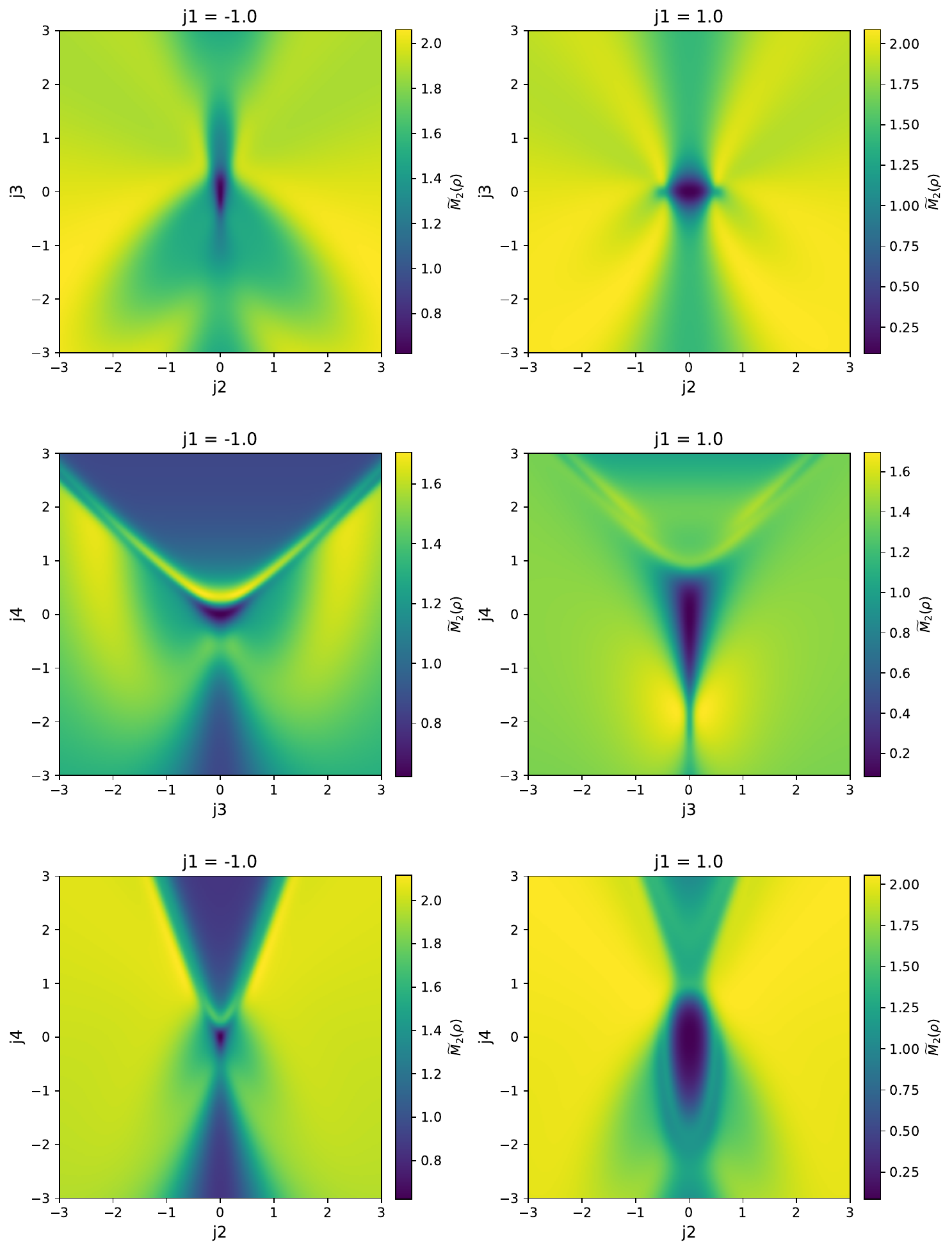}
    \caption{Mixed-state stabilizer R\'enyi diagnostic $\widetilde M_2(\rho)$ at $T=0.1$. Left column $\mathcal{J}_1=-1$, right column $\mathcal{J}_1=+1$; rows scan the $(\mathcal{J}_2,\mathcal{J}_3)$, $(\mathcal{J}_3,\mathcal{J}_4)$ and $(\mathcal{J}_2,\mathcal{J}_4)$ planes over $[-3,3]$. Each panel has its own colour scale.}
    \label{fig:magic_1}
\end{figure}
 
The right column (${\mathcal{J}}_1=-1$) is where the two measures part company, and at matched $T$ the
contrast is unambiguous. The concurrence was bright almost everywhere on this side, vanishing only at
the frustrated point. The negativity instead shows a broad dark plume 
fanning downward from the
origin into ${\mathcal{J}}_3<0$ (and analogous dark valleys in the other two planes). This is not a
single-state effect (away from the origin the ground state is a non-dgenerate entangled $2$-in-$2$-out state
with $\bar N\approx C\approx1.3$) but a consequence of the thermal admixture of the remaining
near-degenerate $2$-in-$2$-out states. Because these states carry their partial-transpose negativity
in different local bases, mixing them largely restores positivity of $\rho^{T_A}$ and collapses
$\bar N$, whereas $C_s$, being a weighted average of the (individually large) eigenstate
concurrences, is left almost unchanged. The negativity therefore resolves the near-degeneracy of the
spin-ice manifold that $C_s$ hides 
everywhere except the exact degenerate point.

\begin{figure}[htbp]
    \centering
    \includegraphics[width=80mm]{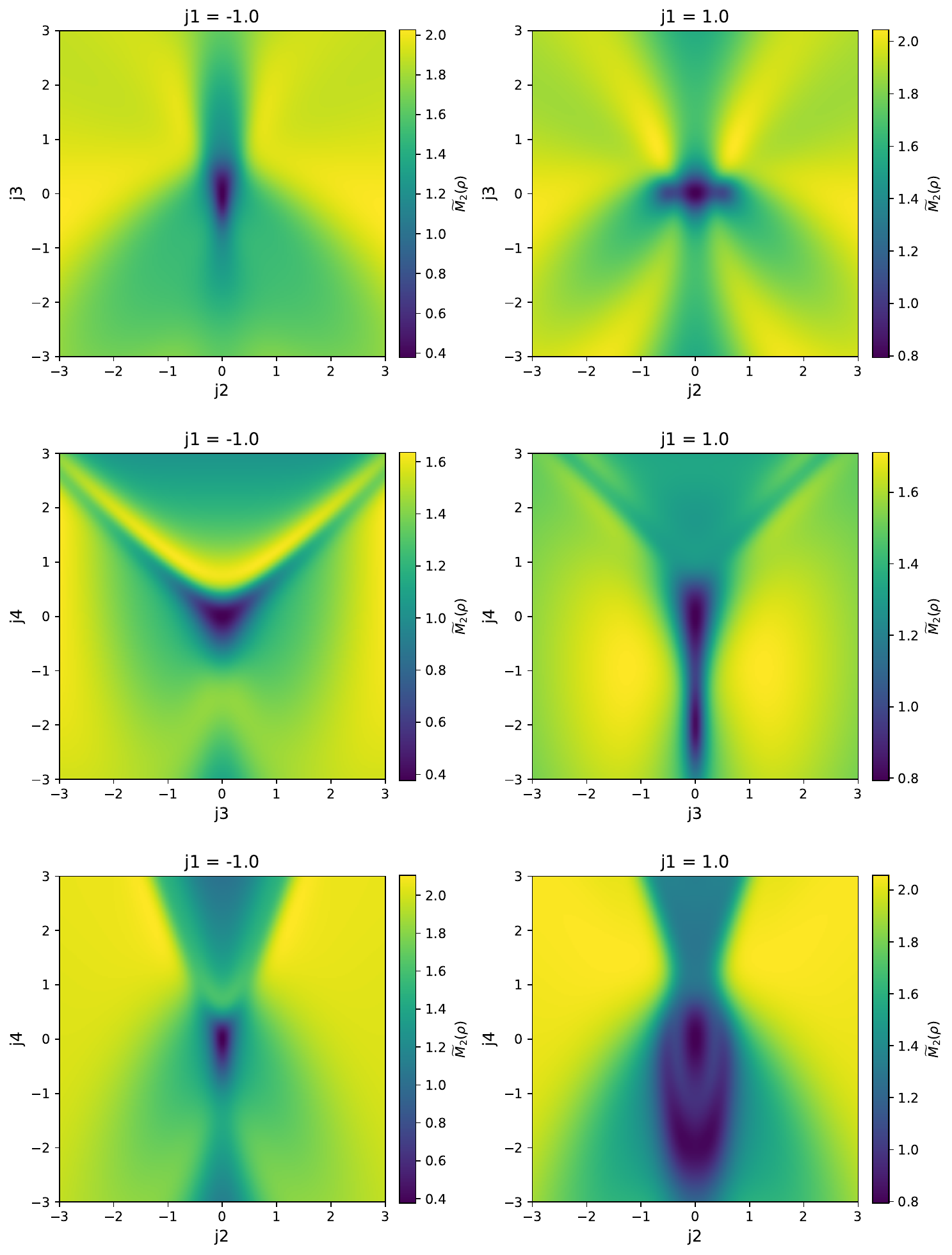}
    \caption{Mixed-state stabilizer R\'enyi diagnostic $\widetilde M_2(\rho)$ at $T=0.2$. Left column $\mathcal{J}_1=-1$, right column $\mathcal{J}_1=+1$; rows scan the $(\mathcal{J}_2,\mathcal{J}_3)$, $(\mathcal{J}_3,\mathcal{J}_4)$ and $(\mathcal{J}_2,\mathcal{J}_4)$ planes over $[-3,3]$. Each panel has its own colour scale.}
    \label{fig:magic_3}
\end{figure}

\paragraph*{Temperature dependence (\cref{fig:negativity_2,fig:negativity_3}).}
The overall scale falls rapidly with temperature: the colour-bar ceilings drop from
$\bar N\lesssim1.2$ at $T=0.1$ to $\sim0.85$ at $T=0.2$ and to $\sim0.1$--$0.45$ at $T=0.5$ (the
${\mathcal{J}}_1=-1$, $({\mathcal{J}}_3,{\mathcal{J}}_4)$ panel peaks at only $\bar N\approx0.10$). This is the
expected approach to $\rho\to\mathbb{1}/16$, which is separable and hence PPT, so $\bar N\to0$; by
$T=0.5$ the negativity is nearly exhausted, surviving only as thin bright rims at large
$|{\mathcal{J}}_a|$ where a single gapped eigenstate still dominates. Over the same window the optimized
concurrence retained order-unity bright regions ($C_s\approx1.2$).
 
\paragraph*{Direct comparison.}
The two families of maps bracket the true mixed-state entanglement from opposite sides. $C_s$ is a
\emph{restricted} convex roof (it minimises only over the symmetric separable densities and so
over-estimates, effectively tracking the entanglement of the thermally populated eigenstates; it
stays large as long as entangled eigenstates carry appreciable Boltzmann weight. $\bar N$ is a PPT
witness on $\rho$ itself) it can vanish on entangled mixtures and so under-reports, but it does
register the decoherence produced by mixing, and therefore decays quickly with $T$. Where a single
pure eigenstate governs the state (low $T$, or the AIAO product region) the two agree in the location
of their nulls \emph{and} in magnitude, as the $A_1$ estimate above anticipates. They diverge in two
regimes: inside the frustrated ${\mathcal{J}}_1=-1$ manifold, where near-degeneracy suppresses $\bar N$
but not $C_s$; and at finite temperature generally, where $\bar N\to0$ while $C_s$ stays finite. The
widening gap between them with increasing $T$ is itself the physical message: the genuine mixed-state
entanglement of the tetrahedron is destroyed by thermal mixing well before the entangled eigenstate
\emph{content} disappears, so a faithful finite-$T$ statement about the spin-ice$\to$entanglement
crossover is better read from $\bar N$ (or the true convex roof) than from $C_s$ alone.

\begin{figure}[htbp]
    \centering
    \includegraphics[width=80mm]{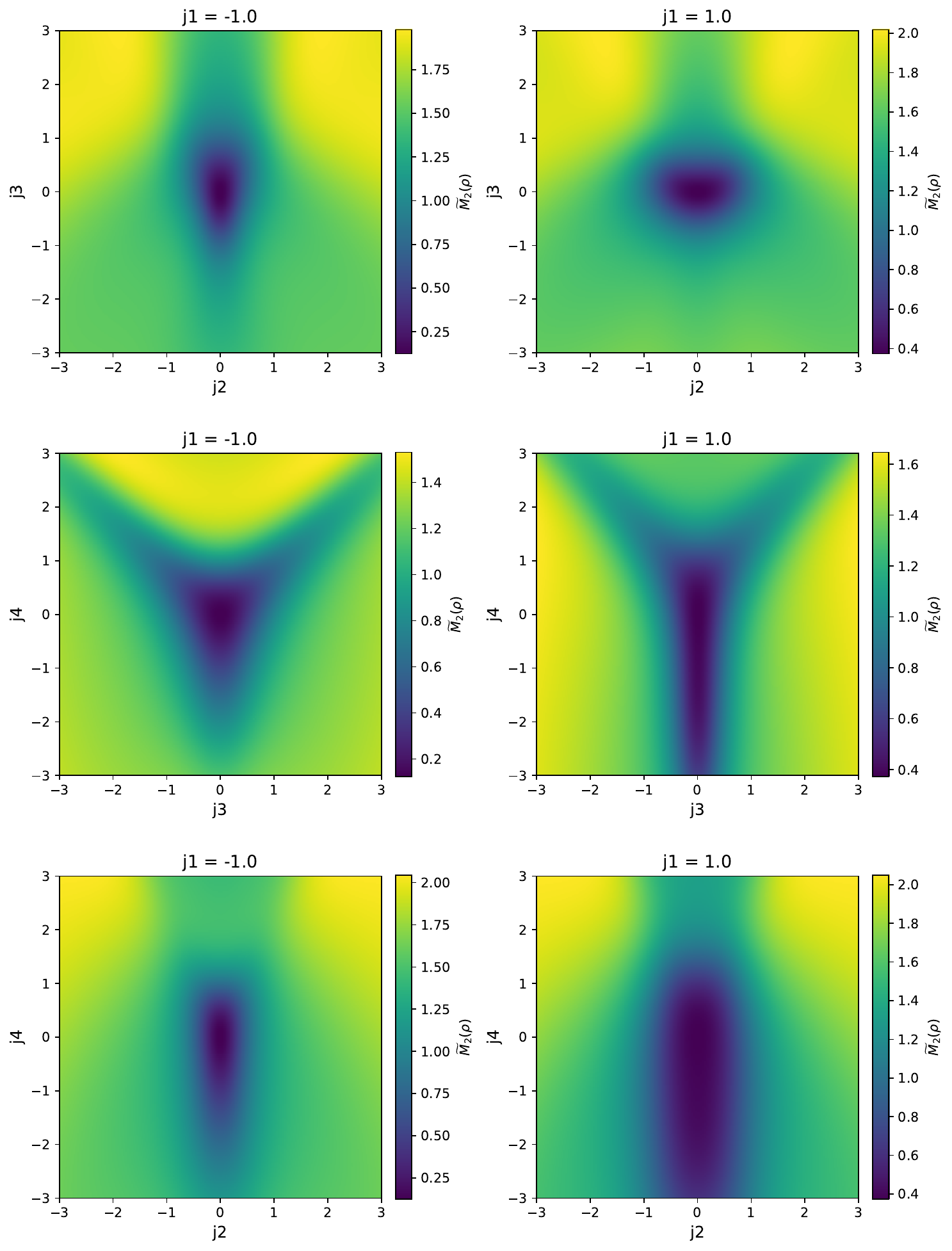}
    \caption{Mixed-state stabilizer R\'enyi diagnostic $\widetilde M_2(\rho)$ at $T=0.5$. Left column $\mathcal{J}_1=-1$, right column $\mathcal{J}_1=+1$; rows scan the $(\mathcal{J}_2,\mathcal{J}_3)$, $(\mathcal{J}_3,\mathcal{J}_4)$ and $(\mathcal{J}_2,\mathcal{J}_4)$ planes over $[-3,3]$. Each panel has its own colour scale.}
    \label{fig:magic_4}
\end{figure}

\subsection{Thermal mixing and stabilizer structure}
\label{sec:thermal-stabilizer}

\begin{figure}[htbp]
    \centering
    \includegraphics[width=90mm]{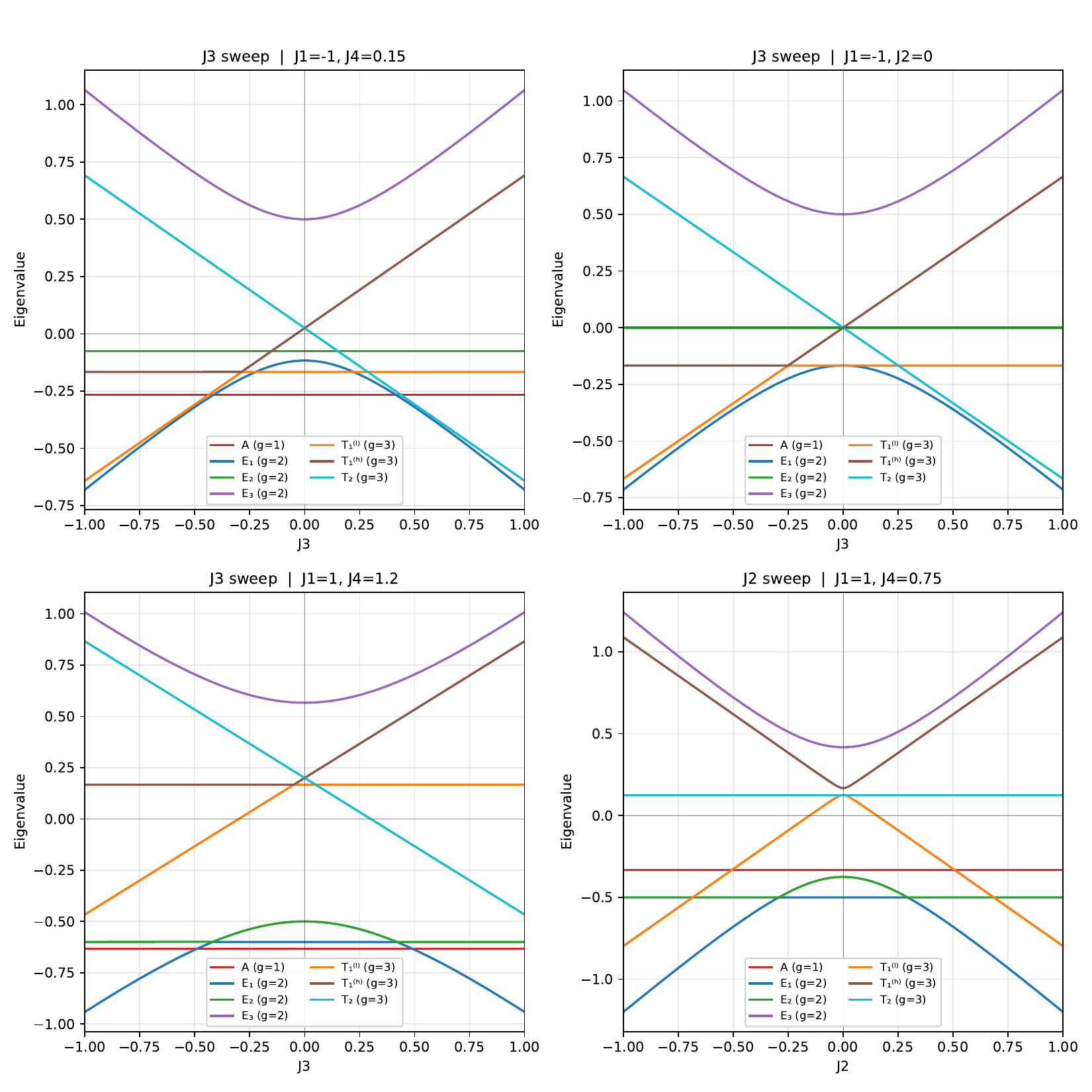}
    \caption{Energy spectra along four exchange-parameter cuts. Top row: $\mathcal{J}_3$ scans for $\mathcal{J}_1=-1$, with $\mathcal{J}_4=0.15$ (left) and $\mathcal{J}_2=0$ (right). Bottom left: a $\mathcal{J}_3$ scan for $\mathcal{J}_1=+1$, $\mathcal{J}_4=1.2$.
Bottom right: a $\mathcal{J}_2$ scan for $\mathcal{J}_1=+1$, $\mathcal{J}_4=0.75$.
The legends identify the symmetry multiplets and their degeneracies $g$.}
    \label{fig:eigenval}
\end{figure}

\begin{figure}[htbp]
    \centering
    \includegraphics[width=80mm]{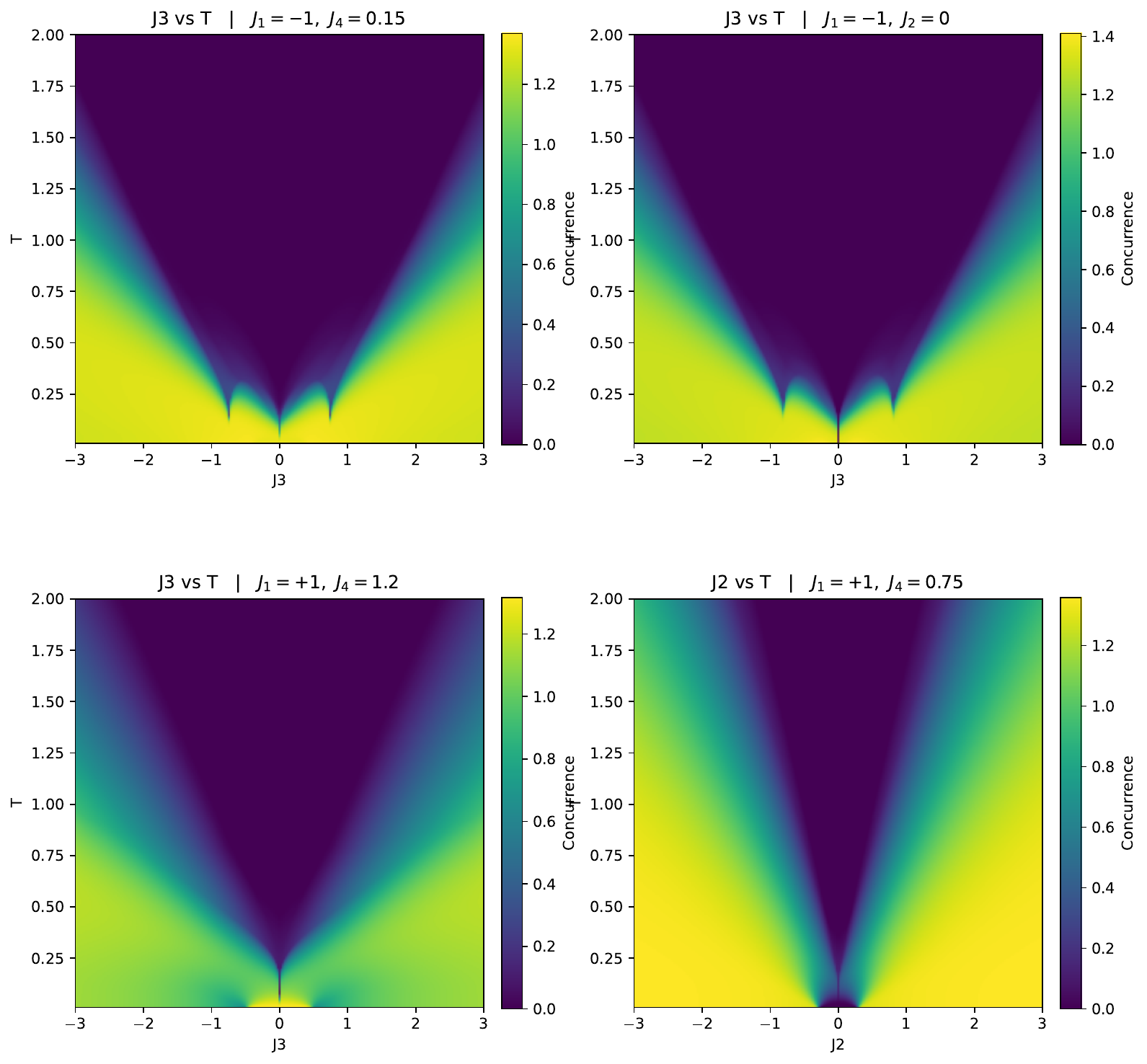}
    \caption{Optimized concurrence bound $C_s$ as a function of temperature and exchange coupling for the four cuts in Fig.~\ref{fig:eigenval}, with the same panel order and fixed parameters. The horizontal axis is $\mathcal{J}_3$ in the first three panels and $\mathcal{J}_2$ in the bottom-right panel, with the scanned range extended to $[-3,3]$.}
    \label{fig:JvsT}
\end{figure}

The spectra in \cref{fig:eigenval} show that the four cuts
have different low-energy structures. For $\mathcal{J}_1=-1$
and $\mathcal{J}_4=0.15$, the singlet $A$ is separated from the
other levels near $\mathcal{J}_3=0$, whereas the second
$\mathcal{J}_1=-1$ cut has a degenerate lowest level there.
At zero swept coupling, the two $\mathcal{J}_1=+1$ cuts instead
have singlet and doublet ground levels, respectively. These
distinctions matter because the thermal density includes all
partners of each populated multiplet. In \cref{fig:JvsT},
the region of small $C_s$ broadens with temperature around
zero swept coupling, while nonzero values extend to higher
temperatures at larger coupling magnitudes. The additional
indentations in the $\mathcal{J}_1=-1$ panels show that this
dependence is not uniform. Their detailed origin requires
the eigenstates and the optimized decomposition as well as
the energy spectrum.

\begin{figure}[htbp]
    \centering
    \includegraphics[width=80mm]{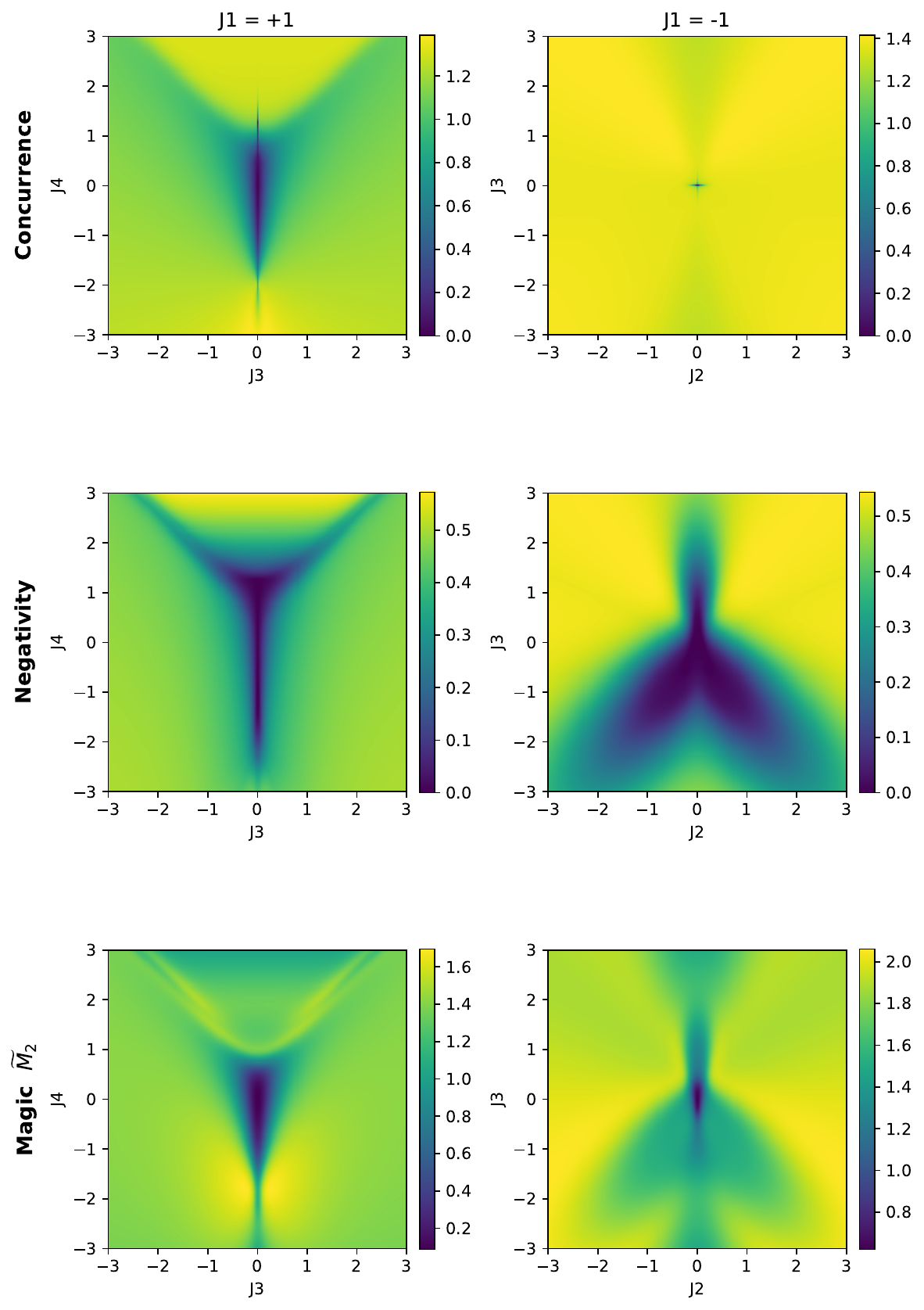}
    \caption{Comparison of the optimized concurrence bound $C_s$ (top row), average negativity $\bar N$ (middle row), and mixed-state stabilizer R\'enyi diagnostic $\widetilde M_2$ (bottom row) at $T=0.1$. Left column: $\mathcal{J}_1=+1$ in the $(\mathcal{J}_3,\mathcal{J}_4)$ plane. Right column: $\mathcal{J}_1=-1$ in the $(\mathcal{J}_2,\mathcal{J}_3)$ plane. Both scanned couplings range from $-3$ to $3$; each panel has its own colour scale.}
    \label{fig:conc-neg-magic}
\end{figure}

\begin{figure}[htbp]
    \centering
    \includegraphics[width=80mm]{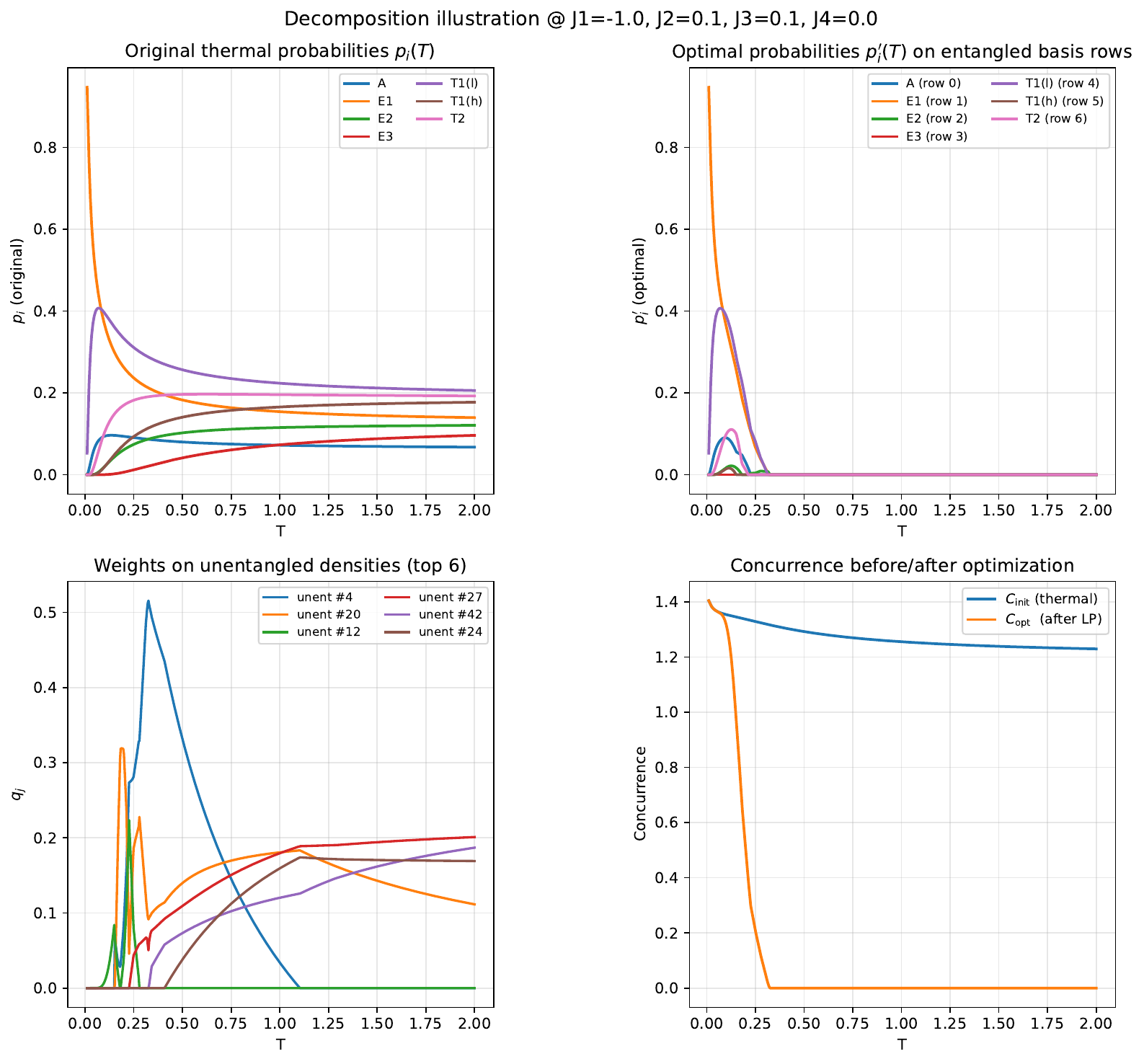}
    \caption{Temperature dependence of the thermal-state decomposition at $\mathcal{J}_1=-1$,
$\mathcal{J}_2=\mathcal{J}_3=0.1$, and $\mathcal{J}_4=0$. Top left: original thermal multiplet weights $p_i$. Top right: optimized weights $p_i'$ on the entangled components. Bottom left: weights $q_l$ of six selected separable densities. Bottom right: the unoptimized eigenstate
concurrence average and the optimized bound $C_s$, which reaches zero near $T\simeq0.3$.}
    \label{fig:decomp}
\end{figure}

\cref{fig:decomp} illustrates the role of the
decomposition at
$(\mathcal{J}_1,\mathcal{J}_2,\mathcal{J}_3,\mathcal{J}_4)
=(-1,0.1,0.1,0)$.
The original thermal weights remain finite as excited
multiplets become populated. In contrast, the optimized
weights $p_i'$ on the entangled components disappear near
$T\simeq0.3$, with the density represented entirely by
separable components. The optimized bound consequently
reaches the plotted zero, although the unoptimized eigenstate
average remains large and is still about $1.2$ at $T=2$.
This is a change in the representation of the same thermal
density; the optimization does not alter its physical
populations.

Subject to numerical accuracy, a vanishing bound
supplies a separable decomposition. A positive bound alone
does not establish entanglement.

The mixed-state stabilizer R\'enyi diagnostic,
\begin{equation}
\widetilde M_2(\rho)=-\log_2\!\left[
\frac{\sum_P[\operatorname{Tr}(\rho P)]^4}
     {\sum_P[\operatorname{Tr}(\rho P)]^2}\right],
\label{eq:sre-results}
\end{equation}
where $P$ runs over the four-spin Pauli strings, displays
a different temperature dependence in
~\cref{fig:magic_1,fig:magic_3,fig:magic_4}. Values of order two remain in parts
of the $\mathcal{J}_2$ scans at $T=0.5$, while the central
depressions broaden and narrow features become smoother.
Heating does not simply reduce the signal everywhere.
For $\mathcal{J}_1=-1$, the ridge in the
$(\mathcal{J}_3,\mathcal{J}_4)$ plane moves towards larger
positive $\mathcal{J}_4$ as temperature increases, and some
regions acquire a larger signal. The separate colour scales
must therefore be taken into account when comparing
temperatures.

The comparison at $T=0.1$ in \cref{fig:conc-neg-magic}
makes the distinction between the diagnostics particularly
clear. For $\mathcal{J}_1=+1$, all three show a central
depression near $\mathcal{J}_3=0$, but the upper central
region at positive $\mathcal{J}_4$ has large concurrence
and negativity without a corresponding maximum of
$\widetilde M_2$. For $\mathcal{J}_1=-1$, the concurrence
map is comparatively featureless away from the origin,
whereas negativity and $\widetilde M_2$ both resolve a
vertical depression near $\mathcal{J}_2=0$ and a broader
structure extending into $\mathcal{J}_3<0$. Their similar
shapes do not imply equivalent measures or coincident zeros.

These results separate the entanglement carried by individual
eigenstates from the properties of their thermal mixture.
They also identify coupling regions in which the different
diagnostics respond differently to heating. For the mixed-state
expression in ~\cref{eq:sre-results}, however, a finite value
can arise even from a mixture of stabilizer states. Its
persistence therefore does not by itself demonstrate usable
magic. Likewise, vanishing negativity does not generally
establish separability. The comparison provides a basis for
investigating thermal robustness, while retaining these
distinctions between an entanglement bound, an entanglement
test and a diagnostic of Pauli expectation values.

\section{Conclusions}
\label{sec:conclusions}

We have used tetrahedral symmetry to investigate multipartite concurrence in the thermal state of the general four-parameter spin-$1/2$ exchange model. Organizing the sixteen states into seven symmetry multiplets reduces the restricted decomposition problem to constrained linear optimization. This provides a practical upper bound $C_s$ on concurrence and, wherever the bound vanishes, an explicit demonstration of full separability.

The concurrence maps reveal two distinct low-temperature regimes. For $\mathcal{J}_1=+1$, the all-in--all-out limit is surrounded by extended regions with $C_s=0$, whose lens and lobe shapes depend strongly on which exchange couplings are varied. For $\mathcal{J}_1=-1$, values of $C_s\simeq1.3$--$1.4$ occupy much of the scanned parameter space, with suppression concentrated near the spin-ice point (\cref{fig:concurrence_4}). The extent and shape of the regions with vanishing concurrence thus distinguish the two limits and their response to the different exchange terms. 

These regimes also clarify the role of eigenstate entanglement in a degenerate system. The all-in--all-out ground manifold has a product-state basis, whereas the spin-ice manifold admits entangled symmetry eigenstates whose equal mixture is separable. Additional exchange interactions can lift the spin-ice degeneracy and select entangled low-energy states. For a nondegenerate ground state, $C_s$ approaches its pure-state concurrence as $T\to0$; residual degeneracy instead requires the mixed density to be considered even in this limit. The symmetry-guided construction therefore connects the concurrence maps to the distinction between entangled eigenstates and entanglement of their statistical mixture. 

The temperature dependence shows how this distinction becomes important away from the ground-state limit. Heating broadens the region of small $C_s$ around the spin-ice point, while nonzero bounds persist to higher temperatures at larger coupling magnitudes (\cref{fig:concurrence_11,fig:JvsT}). Thermal occupation of nearby multiplets can consequently change the result substantially even when their individual concurrences are large. At $\mathcal{J}_1=-1$, $\mathcal{J}_2=\mathcal{J}_3=0.1$, and $\mathcal{J}_4=0$, the optimization finds a fully separable representation near $T\simeq0.3$, within numerical accuracy, although the unoptimized eigenstate average remains about $1.2$ at $T=2$ (~\cref{fig:decomp}). This is a concrete gain from the concurrence optimization: it establishes separability by constructing the required decomposition, which a vanishing negativity alone cannot do.

The negativity and the stabilizer R\'enyi diagnostic complement these concurrence results. Nonzero negativity confirms entanglement in parts of the parameter space at $T=0.5$; a positive upper bound $C_s$ alone cannot provide that confirmation. The comparison with $\widetilde M_2$ shows that its maxima need not coincide with large concurrence bounds or negativity. Regions combining nonzero negativity with large $\widetilde M_2$ identify candidates for studying entanglement and magic together, although this mixed-state diagnostic does not by itself certify usable magic.

The main outcome is a symmetry-based account of how exchange interactions select entangled states and how thermal mixing can restore separability. The concurrence maps and optimized decompositions make these competing effects explicit within one frustrated spin cluster. Together with verified finite-temperature entanglement, they provide a practical basis for choosing couplings at a specified temperature and for extending the study of entanglement to coupled tetrahedra.

\section*{Acknowledgments}

This work was funded by the Natural Sciences and Engineering Research Council of Canada, grant number RGPIN-2020-05615.

\bibliographystyle{unsrt}
\bibliography{citations.bib}


\end{document}